\documentclass[twocolumn,twosided]{IEEEtran}
\usepackage{amsmath,epsf}

\numberwithin{equation}{section} 
\newtheorem{lemma}{Lemma}[section]
 \newtheorem{theorem}[lemma]{Theorem}

 \newtheorem{claim}[lemma]{Claim}

\newtheorem{definition}[lemma]{Definition}

 \newtheorem{ex}[lemma]{Example}
\newenvironment{example}{\begin{ex}}{\hspace*{\fill}$\diamondsuit$\end{ex}}
 \newtheorem{rem}[lemma]{Remark}

\newcommand{\ar}{\rightarrow}
\newcommand{\arp}{\Rightarrow}

\newcommand{\free}{\bot}

\newcommand{\s}[1]{s(#1)}
\newcommand{\f}[1]{f(#1)}

\newcommand{\jump}{\hspace*{1em}}

\newenvironment{prot}{\tt\begin{tabbing}}{\end{tabbing}}

\newenvironment{hist}{\begin{verse}}{\end{verse}}

\begin{document}
\title{Randomized Two-Process Wait-Free Test-and-Set}


\author{John Tromp and Paul Vitanyi\thanks{
J. Tromp is with the
Centrum voor Wiskunde en Informatica,
Kruislaan 413, 1098 SJ Amsterdam, The Netherlands. email: tromp@cwi.nl.
P.M.B. Vit\'anyi is with the Centrum voor Wiskunde en Informatica and
the University of Amsterdam, address: CWI,
Kruislaan 413, 1098 SJ Amsterdam, The Netherlands, email: paulv@cwi.nl.
Both authors were partially supported by the
EU fifth framework project QAIP, IST--1999--11234,
the NoE QUIPROCONE IST--1999--29064,
the ESF QiT Programmme, and the EU Fourth Framework BRA
 NeuroCOLT II Working Group
EP 27150.
}
}

\maketitle

\begin{abstract}
We present the first explicit, and currently simplest,
randomized algorithm for
two-process wait-free test-and-set. It is implemented with two
4-valued single writer single reader atomic variables.
A test-and-set takes at most 11 expected elementary steps, while
a reset takes exactly 1  elementary step.
Based on a finite-state analysis, the proofs of correctness
and expected length are compressed into one table.
\end{abstract}

\section{Introduction}
\label{sect.intro}
A test-and-set protocol concurrently executed by each process out of
a subset of $n$ processes
selects a {\em unique} process from among them.
In a distributed or concurrent system, the test-and-set operation is
useful and sometimes mandatory in a variety of situations including
mutual exclusion, resource allocation,
leader election and choice coordination. It is well-known
that in the wait-free setting,
\cite{La86}, a deterministic construction from atomic read/write
variables is impossible \cite{LA-A87}.  Although widely assumed to
exist, and referred to,
an explicit randomized construction for wait-free
test-and-set has not appeared in print yet, apart from a deterministic
construction {\em assuming} two-process atomic test-and-set \cite{AGTV91}. 
The latter, in the form of a randomized two-process wait-free test-and-set
has been circulated in draft form \cite{TV90} for 
a decade. Here we finally present the construction. 
Since such constructions are notoriously prone to hard-to-detect errors,
we prove it correct by an exhaustive finite-state proof, thus also
presenting a nontrivial application of this proof  technique.

{\bf Interprocess Communication:}
The model is interprocess communication through shared memory as
commonly used in the theory
of distributed algorithms
\cite{Ly96}.
We use atomic single writer single reader registers as primitives. 
Such primitives
can be implemented wait-free
from single-reader single-writer ``safe''
bits 
(mathematical versions of hardware ``flip-flops'')
\cite{La86}). 
A concurrent object is {\em constructible\/}
if it can be implemented deterministically with boundedly many safe bits.
A deterministic protocol executed by $n$ processes is {\em wait-free}
if there is a finite function $f$ such that every non-faulty process terminates
its protocol executing a number of at most $f(n)$ of accesses
to the shared memory primitives, regardless of
the other processes execution speeds.
If the execution speed of a process drops to zero then
this is indistinguishable from the process having a crash failure.
As a consequence, a wait-free solution can tolerate up to
$n-1$ processes having crash failures 
(a property called ``$(n-1)$-resiliency''), since the surviving
non-faulty process correctly executes and terminates its protocol.
Below, we also write ``shared variable'' for ``register.''

{\bf Randomization:}
The algorithms executed by each process are randomized by
having the process flip coins (access a random number generator).
In our randomized algorithms
the answers are always correct---a unique process gets selected---
but with small probability
the protocol takes a long time to finish.
We use the customary assumption
that the coin flip and subsequent
write to shared memory are separate atomic actions.
To express the computational complexity of our algorithm we use
the expected complexity, over all system executions and
with respect to the randomization by the processes and the worst-case
scheduling strategy of an adaptive adversary.
A {\em randomized} protocol is wait-free if $f(n)$
upper bounds the {\em expectation} of the number of elementary steps,
where the expectation is taken over all randomized system
executions against the worst-case adversary in the class
of adversaries considered (in our results the
adaptive adversaries).

{\bf Complexity Measures:}
The computational complexity of distributed
deterministic algorithms using shared memory
is commonly expressed in number and type
of intercommunication primitives required and the maximum number of
sequential read/writes by
any single process in a system execution. Local computation is usually ignored,
including coin-flipping in a randomized algorithm.

{\bf Related Work:}
What concurrent wait-free object is
the most powerful constructible one?
It has been shown that wait-free atomic
multi-user variables, and atomic snapshot objects, 
are constructible,
for example \cite{Pe83,La86,VA86,IL87,Sc88,LTV89,SinAG94,DS89,An90,AADGMS90,HV01}.
In contrast, the agreement problem in
the deterministic model of computation (shared memory or message passing) is
unsolvable in the presence of faults \cite{FisLP,He91a,LA-A87}.
Correspondingly, wait-free consensus---viewed as an object
on which each of $n$ processes can
execute just one operation---is not constructible \cite{CIL87,Ab88},
although randomized implementations are possible
\cite{CIL87,Ab88,As90,SSW90}.
Wait-free concurrent test-and-set can deterministically
implement two-process wait-free consensus, and therefore
is not deterministically constructible \cite{LA-A87,He91a}.
This raises the question of whether randomized algorithms for
test-and-set exist. 

In \cite{He91a} it is shown that repeated use of `consensus' on
{\em unbounded\/} hardware can implement `test-and-set'.
In \cite{Pl89,SSW90,He91} it is argued that a 
bounded solution can be obtained by combining several intermediate
constructions, like so-called ``sticky bits'',
but no explicit construction is presented to back up this claim.  
To quote \cite{Pl89}: ``randomized consensus algorithms of Chor,
Israeli, and Li \cite{CIL87}, Abrahamson \cite{Ab88},
Aspnes and Herlihy \cite{AH}, and Attiya, Dolev, and Shavit \cite{ADS},
together with our construction imply that polynomial number of safe bits is
sufficient to convert a safe implementation into a (randomized)
wait-free one.''
Any such a ``layered'' construction will require
orders of magnitude more primitive building blocks like
one-writer one-reader bits than the direct construction we present below.
Wait-free $n$-process test-and-set can be implemented
deterministically from wait-free two-process test-and-set, \cite{AGTV91},
showing that the impossibility of a deterministic algorithm
for $n$-process test-and-set is solely due to the two-process case.

{\bf Present Results:}
Despite the frequent use of randomized wait-free test-and-set
in the literature, no explicit construction for the basic ingredient, 
randomized wait-free two-process
test-and-set, has appeared in print. Our
construction, \cite{TV90}, has been subsumed and referred to long since, 
for example in \cite{AGTV91,PTTV98,EHW98,BPSV00}, but other
interests prevented us publishing a final version earlier. 
The construction is optimal or close to optimal.
The presented algorithm directly
implements wait-free test-and-set between two processes
from single-writer single-reader atomic shared registers.
Randomization means that the algorithm contains a
branch conditioned on the outcome of a fair coin flip (as in
\cite{Ra82}). 
We use a finite-state based proof technique
for verifying correctness and worst-case expected execution length
in the spirit of \cite{CGP00}.
Our construction is very simple: it uses two 4-valued 1-writer 
1-reader atomic variables.
The worst-case expected number of elementary steps (called ``accesses''
in the remainder of the paper) in a test-and-set operation
is $11$, whereas a reset always takes 1 access.

\section{Preliminaries}
Processes are sequentially executed finite programs
with bounded local variables communicating through
single-writer, multi-reader bounded
wait-free atomic registers (shared variables).
The latter are a common model for interprocess
communication through shared memory as discussed briefly
in Section~\ref{sect.intro}.
For details see \cite{La86,LTV89} and for use and motivation
in distributed protocols see \cite{BarD89,BorG93,HerS93}.

\subsection{Shared Registers, Atomicity}
The basic building blocks of our construction are
4-valued 1-writer
1-reader atomic registers.
Every read/write register is {\em owned} by one process.
Only the owner of a register can write it, while only one other
process can read it. In one {\em access} a process can
either:
\begin{itemize}
\item
Read the value of a register;
\item
Write a  value to one of its own registers;
\item
Moreover, following the read/write of a register
the process possibly flips a local coin (invokes a random number generator
that returns a random bit), preceded or
followed by some local computation.
\end{itemize}

We require the system to be {\em atomic}:
every access of a process can be thought to
take place in an indivisible instance
of time and in every indivisible time instance at most
one access by one process is executed. The atomicity requirement induces
in each actual system execution
total orders on the set of all of
the accesses by the different processes, on the set of accesses
of every individual process, and on the set of
read/write operations executed on each individual register.
The {\em state} of the system gives
for each process:
the contents of the program counter, the contents
of the local variables, and the contents of the owned shared registers.
Since processes execute sequential programs,
in each state every process has at most a single
access to be executed next. Such accesses are
{\em enabled} in that state.

\subsection{Adversary}
There is an {\em adversarial} scheduling demon that in each state
decides which enabled access is executed next, and thus
determines the sequence of accesses of the system execution.
There are two main types of adversaries: the {\em oblivious}
adversary that uses a fixed schedule independent of the
system execution, and the much stronger {\em adaptive}
adversary that dynamically adapts the schedule based on
the past initial segment of the system execution. Our results
hold against the adaptive adversary---the strongest
adversary possible.

\subsection{Complexity}
The computational complexity of a randomized distributed algorithm
in an adversarial setting and the corresponding notion of wait-freeness
require careful definitions. 
For the rigorous
novel formulation of adversaries as restricted measures over the set
of system executions we refer to the Appendix of \cite{PTTV98}.
For the simple application in this paper we can assume that
the notions  of global (system) execution, wait-freeness,
adaptive adversary, and expected complexity are familiar.
A randomized distributed algorithm is {\em wait-free}
if the expected number of read/writes to shared
memory by every participating process is bounded
by a finite function $f(n)$, where $n$ is the number of processes.
The expectation is taken over the probability measure
over all randomized global (system) executions against the worst-case
adaptive adversary.

\section{Test-and-Set Implementation}
We first specify the semantics of the target object:

\begin{definition}\label{def.targetsemantics}
An {\em atomic test-and-set object} $X$ is a {\em global variable},
associated with $n$ processes 
$P_0 ,\ldots, P_{n-1}$, exhibiting the
following functionality:
\begin{itemize}
\item
The value of $X$ is 0 or 1;
\item
Every process $P_i$ has a {\em local binary 
variable} $x_{i}$ which it alone can read or write; 
\item
At any time exactly one
of $X, x_{0} ,\ldots, x_{n-1}$ has value 0, all others
have value 1 (we assume the global time model); 
\item
A process $P_i$ with $x_{i} = 1$
can atomically execute a {\em test-and-set} operation $\tau$:
\begin{hist}
read $x_{i} := X$; write $X := 1$; return $x_{i}$.
\end{hist}
\item
A process~$P_i$ with $x_{i} = 0$ can atomically execute a {\em reset} operation
$\rho$:
\begin{hist}
$x_{i} := 1$; write $X := 0$.
\end{hist}
\end{itemize}
This specification naturally leads to 
the definition of the {\em state\/} of the
test-and-set object as an element of
$\{\free, 0,\ldots,n-1\}$ corresponding to the unique local variable
out of $X, x_{0}, \ldots , x_{n-1}$ that has value 0. Here $\free$ is the state
that none of the $x_i$'s is 0.
Formally, the specification is given later as a finite automaton
in Definition~\ref{def.FA1}.
\end{definition}
Since ``atomicity'' means that the operation is executed
in a single indivisible time instant, and, moreover, in every such
time instant at most one operation execution takes place, 
the effect of a test-and-set operation by process $P_i$ is that
$x_i :=0$ iff all $x_j \neq 0$ for all $j \neq i$, and $x_i =1$ otherwise.
The effect of a reset operation by $P_i$ is only defined for
initially $x_i = 0$ and $x_j \neq 0$ for all $j \neq i$, and results
in $x_i :=1$.
To synthesise the target object from more elementary objects,
we have to use a sequence of atomic accesses to these elementary objects.
By adversary scheduling these sequences may be interleaved arbitrarily.
Yet we would like to have the effect of an atomic execution
of the test-and-set operations and the reset operations by
each process. To achieve such a ``virtual'' atomic execution
we proceed
as follows:

\begin{definition} 
An {\em implementation} of a
{\em test-and-set} operation $\tau$ or a {\em reset} operation $\rho$
by a process $P$ is an algorithm executed by $P$ that results in
an ordered sequence 
of accesses of that process to elements of a set
$\{R_{0}, \ldots ,R_{m-1}\}$ of atomic shared variables,
interspersed with local computation and/or local coin flips. The sequence
of accesses is determined by the,
possibly randomized, algorithm,
and the values returned by the ``read'' accesses to shared variables. 
We denote an access by 
$(P,R,A)$, meaning that process $P$
executes access $A$ (read or write a ``0'' or ``1'')
on shared variable $R$. The implementation must satisfy the
specification of the target test-and-set semantics of Definition
\ref{def.targetsemantics} restricted to process $P$.
Formally, the specification is given later as a finite automaton
in Definition~\ref{def.FA2}.
\end{definition}

\begin{definition}
A {\em local execution} of a process $P$ consists of the (possibly
infinite) sequence
of test-and-set operations and reset operations it executes,
according to the implementation,
each such operation $a \in \{\tau, \rho \}$ 
provided with a {\em start} time $s(a)$
and a {\em finish} time $f(a)$---we assume a global time model.
Note that $s(a)$ coincides with the time of execution of the
first access in the ordered sequence consituting $a$,
 and $f(a)$ coincides the time of execution of
the last access in the ordered sequence
constituting $a$.
By the atomicity of the individual accesses in the global time model,
all accesses are executed at different time instants.
In certain cases (which we show to have zero probability)
it is possible that $f(a)$ is not finite (because the algorithm
executes infinitely many loops with probability $\frac{1}{2}$ each).
\end{definition}

\begin{definition}
Let the local execution of process $P_i$ consist of  
the ordered sequence of operations
$a_1^i , a_2^i , \ldots$ ($0 \leq i \leq n-1$).
A {\em global execution} consists of the $({\cal A}, \rightarrow)$
where ${\cal A} = \{a_j^i : j=1,2, \ldots , \; 0 \leq i \leq n-1 \}$
and $\rightarrow$ is a partial order on the elements of ${\cal A}$
defined by $a \rightarrow b$ iff $f(a) < s(b)$ (the last access of $a$
precedes the first of $b$). We require that
the number of $b$ such that $b \rightarrow a$ is finite for each $a$.
\end{definition}

A test-and-set operation
or reset operation by a particular process may consist
of more than one access, and therefore the
local executions by the different processes may happen concurrently
and asynchronously. This has the effect that
a {\em global execution} can correspond to many 
different interleavings.

\begin{definition}
Consider a global execution.
An {\em interleaving} of the accesses by the different processes associated
with the global execution is
a (possibly infinite) totally ordered 
sequence $(P^1,R^1,A^1), (P^2,R^2,A^2) \ldots$,
where $(P^i,R^i,A^i)$ is the $i$th access,
respecting 
\begin{itemize}
\item
The start times
and finish times determined by the local executions; and
\item
the order of the accesses in the local executions.
\end{itemize}
\end{definition}

The implementation should guarantee that
the functionality of the implementation is ``equivalent'',
in an appropriate sense,
to the functionality of the target test-and-set object, and
in particular satisfies the ``linearizability requirement'' \cite{HW}
(also called ``atomicity'' in \cite{La86}).

\begin{definition}
The system {\em implements} the target test-and-set object if
the system is initially in state $\free$, and
we can extend $\rightarrow$ on ${\cal A}$ to a total 
order $\arp$ on ${\cal A}$ with an initial element, satisfying:
\begin{itemize}
\item From state $\free$, a successful test-and-set operation
$\tau$ executed by process $P_i$ (setting $x_i :=0$)
moves the system
to state $i$ at some time instant in the interval $[s(\tau), f(\tau)]$;
\item from state $i$, a reset operation $\rho$ executed by process
$P_i$  moves the system
to state $\free$ at some time instant in the interval $[s(\rho),f(\rho)]$; 
\item From state $i$, every operation execution different from
a reset by process $P_i$  leaves the system
invariant in state $i$; and
\item No other state transitions than the above are allowed.
\end{itemize}
The implementation must satisfy the
specification of the target test-and-set semantics of Definition
\ref{def.targetsemantics}.
Formally, the specification is given later as a finite automaton
in Definitions~\ref{def.FA3} and \ref{def.FA4}.
\end{definition}

To prove that a protocol
executed by all processes is an implementation of the target test-and-set
object it suffices to show that every possible interleaving that can
be produced by the processes executing the protocol in every global
execution, starting from the $\free$ state, satisfies the above
requirements. 

\section{Algorithm}

\begin{figure*}[htbp]

\centerline{\epsfbox{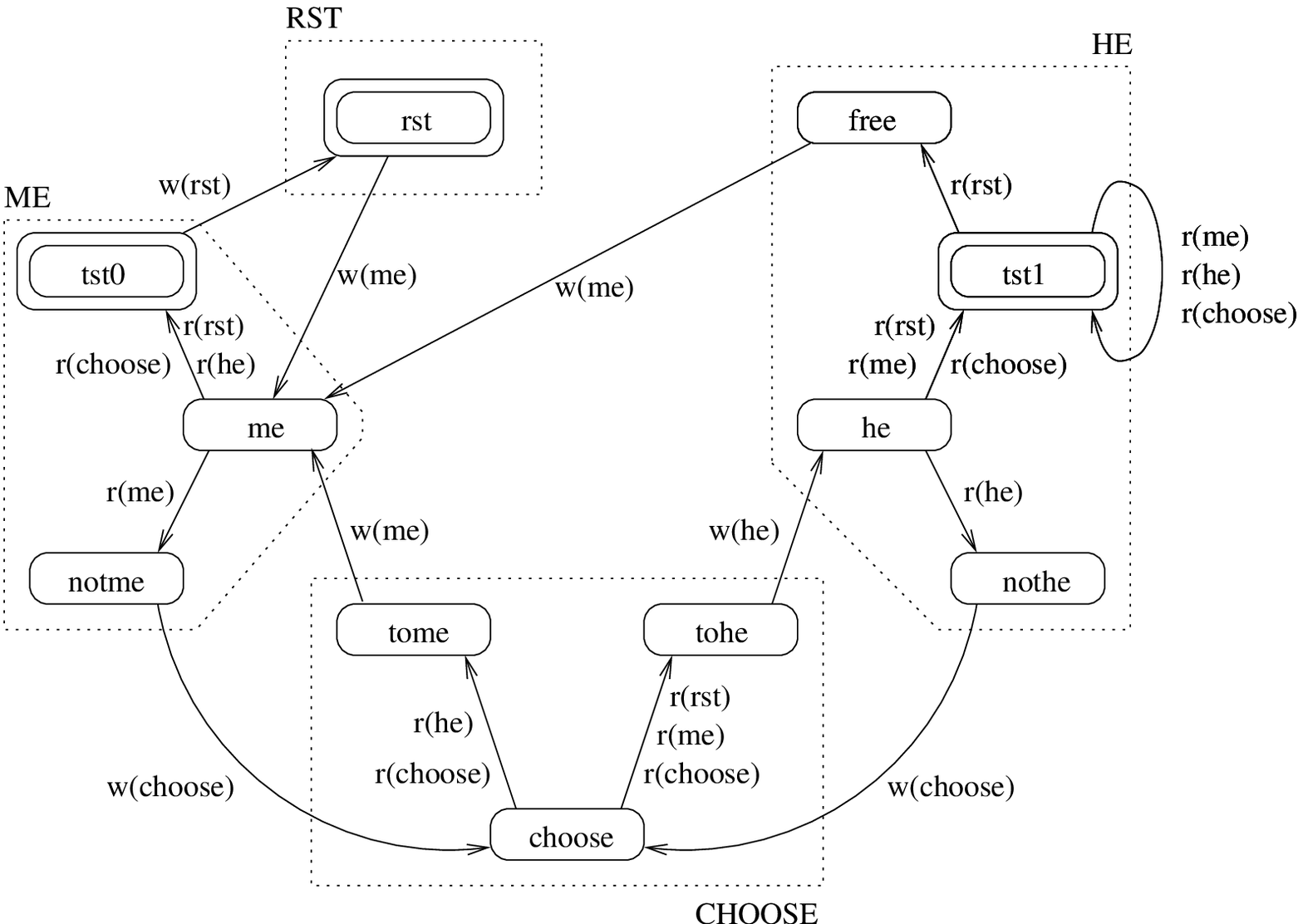}}

\caption{State Chart}
\label{prot2}
\end{figure*}

%
%
We give a test-and-set implementation between two processes,
process~$P_{0}$ and process~$P_{1}$.
The construction uses two 4-valued shared read/write variables
$R_{0}$ and $R_{1}$. The four values are 
`me', `he', `choose', `rst'---chosen
as a mnemonic aid explained below. 
Process~$P_i$ solely writes variable $R_{i}$, its own variable, and
solely reads $R_{1-i}$. 
For this reason the reads and writes in the protocol
don't need to be qualified by the shared variables they access. 
The protocol, for process~$P_i$ ($i=0,1$), is presented
as both a finite state chart, Figure~\ref{prot2} and as
the program below. The state chart representation will simplify
the analysis later. 
The transitions in the state chart are labeled with reads $r(value)$
and writes $w(value)$ of the shared variables, where $value$
denotes the value read or written. 
The 11 states of the state chart are split
into 4 groups enclosed by dotted lines. Each group is
an equivalence class consisting of 
the set of states in which process $P_i$'s own shared variable $R_i$
has the same value. 
That is, the states in a group are equivalent in the sense that
process~$P_{1-i}$ cannot distinguish between them by reading $R_{i}$.
Accordingly, the inter-group transitions are writes to $R_{i}$,
whereas the intra-group transitions are reads of $R_{1-i}$.
Each group is named after the corresponding value of the 
%
%
own shared variable $R_i$. The state chart is deterministic, but for
a coin flip which is modeled by the two inter-group transitions in
the ``choose'' group, representing the two outcomes
of a fair coin flip.
Doubly circled states are ``idle'' states (no
operation execution is in progress), and singly circled states
are intermediate states in an operation execution that is in
progress.

A program representation of the protocol,
for process~$P_i$, is given below.
An occurrence of $R_{i}$ not preceded by `write'
(similarly, $R_{1-i}$ not preceded by `read') as usual refers to the last value
written to it (resp. read from it).
The conditional `rnd(true,false)'
represents the boolean outcome `true' or `false' of a fair coin flip.
The system is initialized with value `rst' in shared variables $R_0,R_1$.
In our protocol, all assignments to local variables consist of 
contents read from
shared variables. To simplify, we abbreviate statements like
``$r_{1-i} := R_{1-i}$; while $r_{1-i}=r_i$ do \ldots $r_{1-i} := R_{1-i}$.'' 
to
``while read $R_{1-i}=R_i$ do \ldots ''.
Here, $r_i$ is the local variable containing the value
last written to shared variable $R_i$ and $r_{1-i}$ is the local
variable storing the last read value of shared variable $R_{1-i}$,
for process $P_i$. This way, our (writing of the) protocol 
can dispense with local variables
altogether.

\begin{prot}
test\_and\_set: \\
\\
if $R_{i}=$ he AND read $R_{1-i}\neq$ rst \\
then return 1 \\
write $R_{i}:=$ me \\
while read $R_{1-i}=R_{i}$ do \\
\jump write $R_{i}:=$ choose \\
\jump if read $R_{1-i}=$ he OR \\
\jump \jump \jump ($R_{1-i}=$ choose AND rnd(true,false)) \\
\jump then write $R_{i}:=$ me \\
\jump else write $R_{i}:=$ he \\
if $R_{i}=$ me \\
then return 0 \\
else return 1 \\
\\
reset: \\
\\
write $R_{i}:=$ rst
\end{prot}

It can be verified in the usual way that the state chart
represents the operation of the program.
The intuition is easily explained
using the state chart.
The default situation is where both processes are idle,
which corresponds to being in the `rst' state.
If process~$P_i$ starts a test-and-set
then it writes $R_{i} :=$ me (indicating its desire to take the 0),
and checks by reading $R_{1-i}$ whether process~$P_{1-i}$ agrees
(by {\em not\/} having $R_{1-i} =$ me).
If so, then $P_i$ has successfully completed a test-and-set by
obtaining the 0 and, implicitly, setting the global variable $X :=1$ . 
In this case process~$P_{1-i}$
cannot get 0 until process~$P_i$ does a reset by writing $R_{i} :=$ rst.
While $R_{i}=$ me, process~$P_{1-i}$ can only move from state `me'
to state `notme' and on via states `choose', `tohe' and `he' to `tst1',
where it completes its test-and-set operation by failure to obtain the 0.

The only complication arises if both processes 
see each other's variable equal to `me'.
In this case they are said to {\em disagree\/} or to be {\em in conflict}.
They then proceed to the `choose' state from where they decide between
going for 0 or 1,
according to what the other process is seen to be doing.
(It is essential that this decision be made in a neutral state,
 without a claim of preference for either 0 or 1. 
If, for example, on seeing a conflict, a process would change preference
at random, then a process cannot know for sure whether the other one agrees
or is about to write a changed preference.)

The deterministic choices, those made if the other's variable is read 
to contain a value different
from `choose', can be seen to lead to a correct resolution
of the conflict. A process ending up in the `tst1' state makes sure
that its test-and-set resulting in obtaining the 1 is justified, 
by remaining in that state
until it can be sure that the other process has taken the 0.
%
%
Only if the other process is seen to be in the `rst' state it
resumes trying to take the 0 itself.

Suppose now that process~$P_i$ has read $R_{1-i}=$ choose and
is about to flip a coin. Assume that process~$1-i$ has already
moved to one of the states `tome'/`tohe' (or else reason with the processes
interchanged).  With 50 percent chance,
process~$P_i$ will move to the opposite state as did process~$P_{1-i}$,
and thus the conflict will be resolved.

In the proof of Theorem~\ref{theo.scci} (below) we establish that
the probability of each loop
through the `choose' state is at most one half, and the expected number
of `choices' (transitions from state choose) is at most two.
This indicates that the worst case expected test-and-set length is 11.
Namely, starting from the `tst1' state,
it takes 4 accesses to get to state `choose', another 4 accesses to loop back
to `choose' and 3 more accesses to reach `tst0'/`tst1'.
The reset operation always takes 1 access.

\section{Proof of Correctness}
The proof idea is as follows: We give a specification of a correct
implementation of two-process test-and-set in the form of a
finite automaton (Figure~\ref{hybrid}). We then show that
all initial segments of
every possible interleaving of accesses by two processes
$P_0$ and $P_1$, both executing the algorithm of the state chart
(Figure~\ref{prot2}), are accepted by the finite automaton.
Moreover, the sequence of states of the finite automaton
in the acceptance process induces a linear order on the 
operation execution of the implemented processes that extends
the partial order induced by the start and finish times of the
individual operation executions. Thus, the implementation is
both correct and atomic. Essentially, the proof is given
by Figure~\ref{tabletje}, which gives the state of the specification finite
automaton for every reachable combination of states which
processes $P_0$ and $P_1$ can attain in their respective copies
of the state chart
(Figure~\ref{prot2}). By analysis of the state chart,
or Figure~\ref{tabletje}, we upper bound the expectation of the number 
of accesses of every operation execution of the implementation by
a small constant.
Hence the implementation is wait-free.  

Let $h$ be an interleaving corresponding to a 
global execution $({\cal A},\ar)$ of two processes running the protocol
starting from the initial state.
Let $\{s(a),f(a): a \in {\cal A}  \}$
be the set of time instants that start or finish an operation execution,
each such time instant corresponding to an access $(P,R,A)$.
Let $B$ denote the set these accesses. 
Recall that if $a$ is a reset, then we have $s(a)=f(a)$ and there is
but a single access executing this operation.

By definition,  $h|B$, the restriction of $h$ to the accesses in $B$,
completely determines the partial order $\ar$. If, for 
every $a \in A$ we can choose a  single access $(P,R,A)_a$ in the sequence
of accesses constituting the operation execution of $a$, such that
if $a \rightarrow b$ then $(P,R,A)_a$ precedes $(P,R,A)_b$ in $h$,
then we are done. Namely, we can imagine an operation $a$
as executing atomically at the time instant of atomic access $(P,R,A)_a$,
and the total order $\arp$ defined by $a \arp b$ iff $(P,R,A)_a$
precedes $(P,R,A)_b$ in $h$, extends the partial order $\ar$.
Denote the set $\{(P,R,A)_a : a \in {\cal A}\}$ by $C$.
We have to show that for every $h$ as defined above such a $C$ can be found.

\begin{definition}{\em Specification of two-process atomic test-and-set:}
\label{def.FA1}
The definition of the target atomic test-and-set for two processes, process $P_0$
and process $P_1$, is captured by
finite automaton FA1 in Figure~\ref{atomic}, which accepts all possible
sequences of atomic test-and-set and reset operations (all states final).
The states are labeled with the owner of the 0-bit.
The arcs representing actions of process $P_1$ are labeled,
whereas the non-labeled arcs represent the
corresponding actions of process $P_0$: 
resulting in setting $x_1 := 1$.

\begin{figure}[htbp]

\centerline{\epsfbox{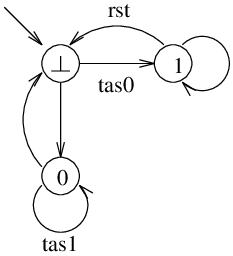}}

\caption{FA1: Specification of two-process atomic test-and-set object}
\label{atomic}
\end{figure}
\end{definition}

\begin{definition}{\em Specification 
of wait-free atomic test-and-set restricted to a single process:}
\label{def.FA2}
Figure \ref{semantics} shows the semantics required of a
correct implementation
of a wait-free test-and-set object as a finite automaton FA2,
that accepts all sequences of accesses by a single process $P_i$
($i=0,1$) executing a correct wait-free atomic test-and-set protocol:
(all states final):
\begin{itemize}
\item the access starting a test-and-set operation execution, denoted s(tas),
\item the atomic occurrence of a test-and-set operation execution
 returning 0, denoted tas0,
\item the atomic occurrence of a test-and-set  operation execution
returning 1, denoted tas1,
\item the access finishing a test-and-set operation execution
returning 0, denoted f(tas0),
\item the access finishing a test-and-set operation execution 
returning 1, denoted f(tas1),
\item the single access corresponding to a complete  reset 
operation execution, denoted rst.
\end{itemize}
These are the events in $B \cup C$ restricted to a process $P_i$.
The reason for not splitting a reset operation execution
into start, atomic occurrence,
and finish is that it is implemented in our protocol as a single
atomic write where the above three transitions coincide.
As before, doubly circled states are ``idle'' states (no
operation execution is in progress), and singly circled states
are intermediate states in an operation execution that is in
progress.

\begin{figure}[htbp]

\centerline{\epsfbox{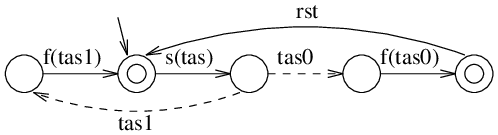}}

\caption{FA2: Specification of 1-process wait-free implementation
of atomic test-and-set}
\label{semantics}
\end{figure}
\end{definition}

\begin{definition}{\em Specification 
of two-process wait-free atomic test-and-set:}
\label{def.FA3}
The proof that our implementation is correct consists in
demonstrating that it satisfies the specification in the form
of the finite automaton FA3
in Figure~\ref{hybrid} below (again all states are final).

\begin{figure}[htbp]

\centerline{\epsfbox{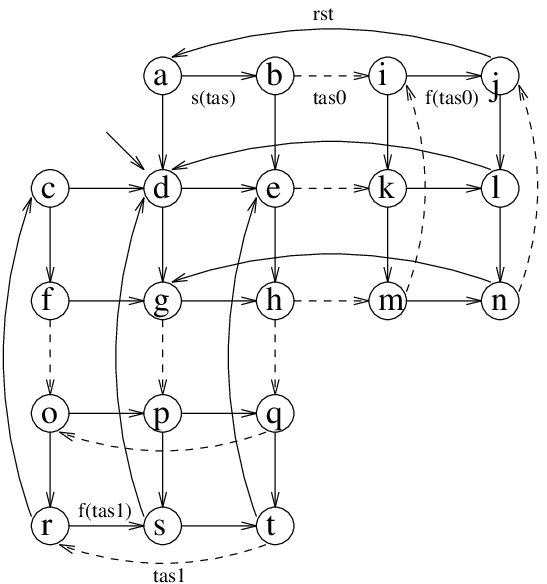}}

\caption{FA3/FA4: Specification of two-process wait-free atomic test-and-set}
\label{hybrid}
\end{figure}

Formally \cite{LT87}, FA3 is the composition of FA1 with two copies of
FA2, in the I/O Automata framework, as follows:
It is drawn as a cartesian product of the two component
processes---transitions of process~$P_0$ are drawn vertically and
those of process~$P_1$ horizontally. For clarity, the transition names
are only given once: only for process~$P_1$.
Identifying the starts and finishes of test-and-set operation executions $a$
with their atomic occurrence  $(P,R,A)_a$ by
collapsing the $\s$ and $\f$ arcs,
FA3 reduces to the atomic test-and-set diagram FA1.
Identifying all nodes in the same column (row) reduces FA3
to FA2 of process~{$P_0$} (process~{$P_1$}).

In the states labeled `a' through `h', neither process owns the $0$;
the system is in state $\free$. 
In the states labeled `i' through `n', process~$1$ owns the $0$;
the system is in state $1$. 
In the states labeled `o' through `t', process~$0$ owns the $0$;
and the system is in state $0$. 
\end{definition}

The broken transitions of
Figure~\ref{hybrid} correspond to the access 
$(P,R,A)_a \in C$, required for a correct implementation,
where the atomic execution of operation $a$  
can be virtually situated. Recall that this is only relevant
for $a$ is a test-and-set operation, since the reset operation
is implemented in the protocol already in a single atomic access
of a shared primitive variable.  

\begin{definition}\label{def.FA4}
Let FA4 be the (nondeterministic) finite automaton
obtained from FA3 by turning the broken transitions of
Figure~\ref{hybrid}, which correspond to the unknown but
existing access $(P,R,A)_a \in C$ where the execution of $a$
can be virtually situated,  
into $\epsilon$-moves.
\end{definition}

\begin{lemma}\label{lem.1}
Acceptance of $h|B$ by FA4 implies that $({\cal A},\ar)$ is linearizable:
partial order $\ar$ can be extended to a total order $\arp$ such
that the sequence of operation executions in ${\cal A}$ ordered
by $\arp$ satisfy the test-and-set semantics specification of
Definition~\ref{def.FA1}.
\end{lemma}

\begin{proof}
If FA4 accepts $h|B$, then, corresponding to the $\epsilon$
moves, we can augment the sequence $h|B$ with an access $(P,R,A)_a$ in
the interval $[s(a),f(a)]$ of each operation 
execution $a \in {\cal A}$---or select the single access involved
if $s(a)=f(a)$ as in the case of a reset operation execution---
to obtain a new sequence $h'$ that is accepted by FA3.
By the way FA1 composes FA3, it accepts $h'|C$, the subsequence of atomic
accesses $(P,R,A)_a$ with $a \in {\cal A}$ contained in $h'$.
Furthermore, letting $t(a)$ denote the time of access $(P,R,A)_a$,
we have  $a \ar b$ iff
$t(a) \leq f(a) \leq s(b) \leq t(b)$. Defining 
$a \arp b$ if $t(a) < t(b)$, the total order of accesses in $h'|C$, 
then $\arp$ is a total order that extends the partial order $\ar$.
That is, the sequence of operation executions of ${\cal A}$,
linear ordered by $\arp$, is accepted by FA1.
\end{proof}

\begin{figure*}[htbp]

\begin{center}
\footnotesize
\begin{tabular}{|c|ccccccccccc|}
\hline
& rst & tst0 & notme & me & tome & choose & tohe & he & nothe & tst1 & free
\\ \hline
rst & d{\bf 10} & l{\bf 10} & cek{\bf 10} & ek{\bf 10} & ek{\bf 10} & c{\bf 10} & c{\bf 10} & c{\bf 10} & c{\bf 10} & d{\bf 10} & ek{\bf 10}
\\
tst0 & s{\bf 1} & * & rt{\bf 1} & rt{\bf 1} & rt{\bf 1} & r{\bf 1} & r{\bf 1} & r{\bf 1} & r{\bf 1} & s{\bf 1} & rt{\bf 1}
\\
notme & agp{\bf 8} & jn{\bf 8} & imoq{\bf 8} & imoq{\bf 8} & * & imoq{\bf 8} & imoq{\bf 8} & o{\bf 4} & * & p{\bf 4} & *
\\
me & gp{\bf 9} & jn{\bf 9} & imoq{\bf 9} & imoq{\bf 9} & imoq{\bf 9} & o{\bf 1} & o{\bf 1} & o{\bf 1} & o{\bf 1} & p{\bf 1} & imoq{\bf 9}
\\
tome & gp{\bf 10} & jn{\bf 10} & * & imoq{\bf 10} & imoq{\bf 10} & imoq{\bf 6} & o{\bf 2} & o{\bf 2} & imoq{\bf 6} & p{\bf 2} & *
\\
choose & a{\bf 3} & j{\bf 3} & imoq{\bf 7} & i{\bf 3} & imoq{\bf 7} & imoq{\bf 7} & imoq{\bf 7} & o{\bf 3} & imoq{\bf 7} & p{\bf 3} & *
\\
tohe & a{\bf 2} & j{\bf 2} & imoq{\bf 6} & i{\bf 2} & i{\bf 2} & imoq{\bf 6} & imoq{\bf 10} & imoq{\bf 10} & * & p{\bf 6} & *
\\
he & a{\bf 1} & j{\bf 1} & i{\bf 1} & i{\bf 1} & i{\bf 1} & i{\bf 1} & imoq{\bf 9} & imoq{\bf 9} & imoq{\bf 9} & p{\bf 5} & *
\\
nothe & a{\bf 4} & j{\bf 4} & * & i{\bf 4} & imoq{\bf 8} & imoq{\bf 8} & * & imoq{\bf 8} & imoq{\bf 8} & p{\bf 4} & *
\\
tst1 & d{\bf 11} & l{\bf 11} & k{\bf 11} & k{\bf 11} & k{\bf 11} & k{\bf 11} & k{\bf 11} & k{\bf 11} & k{\bf 11} & * & *
\\
free & gp{\bf 10} & jn{\bf 10} & * & imoq{\bf 10} & * & * & * & * & * & * & *
\\ \hline
\end{tabular}
\end{center}
\caption{Table verification of correctness and wait-freedom}
\label{tabletje}

\end{figure*}

Recall that Figure~\ref{prot2} is the state chart
of the execution of the implementation
of an operation by a single process. Each process can be in a particular
state of the state chart. Let $(s_0,s_1)$ denote the state
of the system with process $P_i$ in state $s_i$ ($i\in \{0,1\}$).

\begin{definition}
The {\em initial} system state is $(rst,rst)$.
A system state $(s_0,s_1)$ is {\em reachable} from the initial system
state $(rst,rst)$ if there is a sequence $h$ arising from the execution
of our test-and-set implementation, represented by the state chart
of Figure~\ref{prot2}, starting from the initial state 
and ending in state $(s_0,s_1)$.
\end{definition}

\begin{example}\label{ex.reach1}
\rm
In the initial state both processes are in state
`rst'.
Process~$P_0$
can start a test-and-set by executing $w(me)$ and entering
state $me$. 
Suppose process~$P_1$
now starts a test-and-set: it executes $w(me)$ and moves to state
$me$. 
Hence, system states $(me,rst)$ and $(me,me)$ are reachable states.
\end{example}

\begin{definition}
The {\em representative set} of a
reachable system state $(s_{0},s_{1})$ is
a nonempty set $S_{s_{0},s_{1}}$ of FA3/FA4 states, 
as in Figure~\ref{hybrid}, such that:
For every sequence of accesses $h$ starting in the 
initial state and ending in state $(s_{0},s_{1})$,
the set $S_{s_{0},s_{1}}$ is
the set of states in which FA4 can be after processing $h|B$,
excluding those states that have outgoing moves that are $\epsilon$-moves only.
\end{definition}

\begin{example}\label{ex.reach2}
\rm
We elaborate Example~\ref{ex.reach1}.
In the initial state both processes are in state
`rst'.
The corresponding start state $d$ of FA4 gives the associated
(in this case singleton) representative set $\{d\}$.
When process~$P_0$
executes $w(me)$ and enters
state $me$, the resulting system state is $(me,rst)$
with the associated representative set $\{g,p\}$ of FA4 states.
That is, the system is now either in state $g$,
meaning that process~$P_0$ has executed $s(tas)$, or in state $p$
meaning that process~$P_0$ has executed $s(tas)$ and also
$tas0$ atomically.
In the scenario of Example~\ref{ex.reach1}, process~$P_1$
now executes $w(me)$ and moves to state
$me$, resulting in the system state $(me,me)$.
The corresponding representative set of FA4 states is $\{i,m,o,q\}$.
State $m$ says process~$P_1$ has executed
$s(tas)$ and $tas0$ atomically, while process~$P_0$
has only executed $s(tas)$---hence the system was previously in state $g$
and not in state $p$. State $i$ says process~$P_1$ has executed
$s(tas)$ and $tas0$ atomically, while process~$P_0$
has executed $s(tas)$ and $tas1$
atomically---and hence the system was previously in state $g$
and not state $p$. States $o$ and $q$ imply the same state
of affairs with the roles of process~$P_0$ and process~$P_1$
interchanged, and the previous system state is
either $p$ or $g$.
(The correspondence between reachable states and their representative
sets is exhaustively established in Claim~\ref{claim.1} below.)
\end{example}

\begin{lemma}\label{lem.2}
Let $h$ be a sequence of accesses arising from the execution
of our test-and-set implementation, represented by the state chart
of Figure~\ref{prot2}, starting from the initial state
(both processes in state `rst'). Then, every initial segment of
$h|B$ is accepted by FA4 starting from initial state `d'. 
\end{lemma}

\begin{proof}
We show that the set of letters in an entry in the table 
of Figure~\ref{tabletje} 
is a representative set
for the state of process $P_0$,  indexing the row, and the
state of process $P_1$, indexing the column.
The entries were chosen excluding all states from the representative
sets with all outgoing moves consisting of $\epsilon$-moves
(but the representative sets contain the
states the outgoing $\epsilon$-moves of the excluded states point to).
This gives the most insight
into the workings of the protocol by considering only the result
of executing $\epsilon$-moves from a state 
if its only outgoing moves are $\epsilon$-moves. 
A $*$-entry indicates an unreachable state pair.
(The number ending an entry gives
the expected number of accesses to finish the current
operation execution of process~$P_0$---and by symmetry, that for
an equivalent state pair with respect to $P_1$. We will use this later.)
Thus, every state $(s_0,s_1)$ of the implementation execution
corresponds with a set of states $S_{s_0,s_1}$ of FA4.

\begin{claim}\label{claim.1}
The representative sets are given by the entries of Figure~\ref{tabletje}.
\end{claim}
\begin{proof}
The proof of the claim is contained
in the combination of Figures~\ref{prot2}, \ref{hybrid}, \ref{tabletje}.
Below we give the inductive argument. The mechanical verification
of the subcases has been done by hand, and again by machine. 
The setting up of the exhaustive list subcases and subsequent verification
by a computer program is the essennce of a finite-state proof. In
this particular case, exceptionally, the finite state machines involved (and
the table of representative sets) have been minimized so that
``mechanical'' verification by hand by the reader is still feasible.
Induction is on the length of the sequence of accesses:
\begin{description}
\item
{\em Base Case:}
Initially, after an empty sequence of accesses, FA4 is in
the state
$\{d\} = S_{rst,rst}$.
\item
{\em Induction:}
Every non-reachable state has a $*$-entry in the table of Figure \ref{tabletje}.
Consider an arbitrary atomic transition from a reachable state $(s_{0},s_{1})$
to a state $(t_{0},t_{1})$, that is, using a single arc in the state chart in 
Figure~\ref{prot2} for either process $P_0$ or $P_1$.
This way, either  $t_0 = s_0$ or
$t_1 = s_1$ but not both.  Then, for every FA4 state $y \in S_{t_{0},t_{1}}$,
Figure~\ref{hybrid},
according to the table of Figure~\ref{tabletje},
there is an FA4 state $x \in S_{s_{0},s_{1}}$ according to 
Figure~\ref{hybrid}, such that 
FA4 can move from $x$ to $y$ by executing: either the access corresponding
to the transition in the state chart in Figure~\ref{prot2},
if that access belongs to $B$, 
or no access otherwise (there is
a sequence of $\epsilon$-moves from $x$ to $y$).
\end{description}

\end{proof}
Since every reachable state of the system $(s_0,s_1)$, with $s_i$ 
($i \in \{0,1\})$ a state of the state chart of Figure~\ref{prot2},
has a representative set in FA4, Figure~\ref{hybrid}, and every state of 
of FA4 is an accepting state, the lemma follows from Claim~\ref{claim.1}.
\end{proof}

\begin{theorem}
The algorithm represented by state chart
of Figure~\ref{prot2}
correctly implements an atomic test-and-set object.
\end{theorem}
\begin{proof}
By Lemma \ref{lem.2} the implementation by the state chart in Figure~\ref{prot2}
correctly implements the specification of two-process test-and-set
given by Figure~\ref{hybrid}. The implementation is linearizable
(atomic) by Lemma~\ref{lem.2}. The system makes progress (every
operation execution is executed completely except for possibly
the last one of each process) since $h|B$ contains only the start
and finish accesses of each operation execution performed by
the implementation. 
\end{proof}

\begin{theorem}
\label{theo.scci}
The algorithm represented by state chart
of Figure~\ref{prot2}
is wait-free: the expected number of accesses to shared variables
never exceeds 11 during execution of an operation.
\end{theorem}
\begin{proof}
In Figure~\ref{prot2} every arc is an access. Double circled
states are idle states (in between completing an operation execution
and starting a new one). Consider process $P_0$ (the case
for process $P_1$ is symmetrical). The longest path without
completing an operation and without cycling is from state `tst1':
tst1, free, me, notme, choose, tohe, he, tst1. This takes 7 accesses.
Four of these accesses are parts of a potential cycle of length 4.
The remainder is 3 accesses outside the potential cycle.
In state `choose', 
the outgoing arrow is a random choice only
when process $P_1$ is also in the CHOOSE
group. If it is, then with $\frac{1}{2}$ probability  $P_1$ 
makes (or has already made) a choice which will cause process
$P_0$ to loop back to the `choose' state again. 
This can happen
again and again.
The expected number
of iterations of loops is $\sum_{i=1}^{\infty} 
i \left( \frac{1}{2} \right)^{i} = \frac{1}{2} (1- \frac{1}{2})^{-2} = 2$. 
Since a loop has length 4, this gives a total of expected accesses of 8
for the loops. Together with 3 non-loop accesses
the total is at most $11$ accesses. 
Such a computation holds for every state
in the state chart of Figure~\ref{prot2}, the only loop being 
the one discussed but the longest possible path is the one starting from 
`tst1'. For definiteness,
we have in fact computed the expected number
of accesses for every accessible state $(s_0,s_1)$ according to
the state chart of Figure~\ref{prot2}, and added that number
to the representative set concerned in the table of Figure~\ref{tabletje}.
Since the expected number of accesses is
between 1 and 11 for all operation executions,
the algorithm given by the state chart of Figure~\ref{prot2} is wait-free.
\end{proof}

To aid intuition, we give an example of checking a few transitions below,
as well as giving the interpretation. 

\begin{example}
We elaborate and continue Examples~\ref{ex.reach1}, \ref{ex.reach2}.
In the initial state both processes are in state
`rst'. 
In Figure~\ref{hybrid},
the table entry $d10$ gives the corresponding start state $d$ of FA4.
The worst-case expected number
of accesses for a test-and-set by process~$0$ is 10. Process~$P_0$
can start a test-and-set by executing $w(me)$ and entering
state $me$. The corresponding table entry $gp9$ 
indicates in Figure~\ref{hybrid}
that the system is now either in state $g$ 
meaning that process~$P_0$ has executed $s(tas)$, or in state $p$
meaning that process~$P_0$ has executed $s(tas)$ and also 
$tas0$ atomically. 
The expected number of accesses is now $9 \leq 10-1$. Suppose process~$P_1$
now starts a test-and-set: it executes $w(me)$ and moves to state
$me$. The corresponding table entry $imoq9$ 
gives the system state as one
possibility in $\{i,m,o,q\}$ in Figure~\ref{hybrid}
and the expected number of accesses for execution of test-and-set by
process~$P_0$ is still 9. State $m$ says process~$P_1$ has executed
$s(tas)$ and $tas0$ atomically, while process~$P_0$
has only executed $s(tas)$---hence the system was previously in state $g$
and not in state $p$. State $i$ says process~$P_1$ has executed
$s(tas)$ and $tas0$ atomically, while process~$P_0$
has executed $s(tas)$ and $tas1$
atomically---and hence the system was previously in state $g$
and not state $p$. States $o$ and $q$ imply the same state
of affairs with the roles of process~$P_0$ and process~$P_1$
interchanged, and the previous system state is
either $p$ or $g$. 

Note that at this point the system can also be in state
$h$ of FA4---both processes having executed $s(tas)$
but no process having executed $tas0$ or $tas1$. However, from $h$
there are two $\epsilon$-moves possible, and no other moves,
leading to $q$ and $m$. This corresponds to the fact that if both
processes have executed $s(tas)$, one of them must return 0 and
the other one must return 1.
We have optimized the table
entries by eliminating such spurious intermediate states $h$ with 
outgoing moves that are $\epsilon$-moves only.

Process~$P_0$ might now read $R_{1} = me$, and move via state `notme' 
(table entry $imoq8$) by writing $R_{0} := choose$,
to state `choose'. Process~$P_1$ is idle in the meantime.
The table entry is now $i3$. This says that
process~$P_1$ has atomically executed $tst0$, and process~$P_0$
has atomically executed $tst1$. Namely, all subsequent schedules
lead in 3 accesses of process~$P_0$ to state `tst1'---hence the expectation 3.

The expected number of remaining accesses of process~$P_0$'s test-and-set
has dropped from 8 to 3 by the last access since 8 was the worst-case
which could be forced by the adversary. Namely, from the system
in state $(notme, me)$, the adversary can schedule process~$P_1$
to move to $(notme, notme)$
with table entry $imoq8$, followed by a move of process~$P_1$ to
state $(notme, choose)$ with table entry $imoq8$,
followed by a move of process~$P_0$ to state $(choose,choose)$
with table entry $imoq7$. Suppose the adversary now schedules
process~$P_0$. It now flips a fair coin to obtain the conditional
boolean $rnd(true,false)$. If the outcome is $true$, then
the system moves to state $(tome, choose)$ with entry $imoq6$.
If the outcome is $false$, then the system moves to state
$(tohe, choose)$ with table entry $imoq6$. Given a fair coin,
this access of process~$P_0$ correctly decrements the expected number
of accesses. Suppose the adversary schedules process~$P_1$ in state
$(choose, choose)$. Process~$P_1$ flips a fair coin.
If the outcome is $true$ the system moves to state
$(choose, tome)$ with table entry $imoq7$; if the outcome is $false$
then the system moves to state $(choose, tohe)$ with 
table entry $imoq7$.
\end{example}

\section{Remark on Multi-Process Test-And-Set}

The obvious way to extend the given solution to more than two processes
would be to arrange them at the leaves of a binary tree.
Then, a process wishing to execute an $n$-process test-and-set,
would enter a tournament, as in \cite{PF77},
by executing a separate two-process test-and-set for each node
on the path up to the root.
When one of these fails, it would again descend,
resetting all the tas-bits on which it succeeded, and return 1.
When it succeeds ascending up to the root, it would return 0
and leave the resetting descend to its $n$-process reset.

The intuition behind this tree approach is that if a process $i$ fails
the test-and-set at some node $N$, then another process $j$ will get
to the root successfully and thus justify the value 1 returned by the former.

The worst case expected length of the $n$-process operations is only
$\log n$ (binary logarithm) times more than that of the two-process case.

Unfortunately, this straightforward extension does not work.
The problem is that the other process $j$ need not be the
one responsible for the failure at node $N$, and might have started
its $n$-process test-and-set only after process $i$ completes its own.
Clearly, the resulting history cannot be linearized.

Nonetheless, it turns out that with a somewhat more
complicated construction we can {\em deterministically}
implement $n$-process test-and-set using
two-process test-and-set as primitives \cite{AGTV91}.
This shows that the impossibility of deterministic
wait-free atomic $n$-process test-and-set is completely
due to the impossibility of deterministic wait-free atomic two-process
test-and-set. This latter problem we have just solved by a
simple direct randomized algorithm.


\begin{thebibliography}{9}

\bibitem{Ab88} K. Abrahamson, On achieving consensus using shared memory,
{\em Proc.  7th ACM Symp. Principles of Distributed Computing}, 1988,
291--302.

\bibitem{AADGMS90} Y. Afek, H. Attiya, D. Dolev, E. Gafni, M. Merritt,
N. Shavit, Atomic snapshots of shared memory,
{\em Journal of the ACM}, 40:4(1993), 873-890.

\bibitem{AGTV91} 
Y. Afek, E. Gafni, J. Tromp, P.M.B. Vit\'anyi, Wait-free test-and-set,
pp. 85-94 in: {\em Proc. 6th Workshop on Distributed Algorithms (WDAG-6)},
Lecture Notes in Computer Science, vol. 647, Springer Verlag,
Berlin, 1992.

\bibitem{ADS}
H. Attiya, D. Dolev, N. Shavit, 
Bounded Polynomial Randomized Consensus,
{\em Proc.8th ACM 
Symp. Principles of Distributed Computing}, 1989, 281--293.

\bibitem{An90} J.H. Anderson, Multiwriter composite registers,
{\em Distributed Computing}, 7:4(1994), 175-195.

\bibitem{As90} J. Aspnes, Time- and space-efficient randomized
consensus,
{\em Journal of Algorithms}, 14:3(1993), 414-431.

\bibitem{AH}
J. Aspnes, M. Herlihy,
Fast randomized consensus using shared memory, 
{\em Journal of Algorithms}, 11:3(1990), 441-461, 

\bibitem{BarD89}
A. Bar-Noy and D. Dolev,
A partial equivalence between shared-memory 
and message-passing in an asynchronous fail-stop distributed
environment, {\em Mathematical Systems Theory}, 26(1993), 21--39. 

\bibitem{BorG93}
E. Borowsky and E. Gafni,
Immediate Atomic Snapshots and Fast Renaming,
{\em Proc. 11th ACM Symp. on Principles of Distributed
Computing}, 1992, pp.~41--52.

\bibitem{BPSV00}
H. Buhrman, A. Panconesi, R. Silvestri, and P. Vitanyi
On the importance of having an identity or, is consensus really
Universal?, {\em Distributed Computing Conference (DISC'00)},
Lecture Notes in Computer Science, Vol. 1914,
Springer-Verlag, Berlin, 2000, 134--148.

\bibitem{BJLFP82}
J.E Burns, P.Jackson, N.A. Lynch, M.J. Fischer, G.L. Peterson,
Data Requirements for Implementation of N-process Mutual Exclusion
Using a Single Shared Variable, {\em J. Assoc. Comput. Mach.},
29(1982),183--205.


\bibitem{CIL87} B. Chor, A. Israeli, M. Li,
Wait--Free Consensus Using Asynchronous Hardware,
{\em SIAM J. Comput.}, 23:4(1994), 701--712. 

\bibitem{CGP00}
E. M. Clarke, O. Grumberg, and D. Peled, {\em Model Checking}, MIT Press, 2000.

\bibitem{DS89} D. Dolev and N. Shavit,
Bounded concurrent time-stamp systems are constructible,
{\em Siam J. Comput.},  26(2):418-455, 1997.

\bibitem{EHW98} W. Eberly, L. Higham, J. Warpechowska-Gruca,
Long-lived, fast, waitfree renaming with optimal name space and high throughput,
{\em Proc. 12th Intn'l Distributed Computing Conference (DISC'98)},
Lecture Notes in Computer Science, 1499, 
Springer-Verlag, Berlin, 1998.

\bibitem{HV01}
S. Haldar, P.M.B. Vitanyi, 
Bounded concurrent timestamp systems using vector clocks,
{\em J. Assoc. Comput. Mach.}, 49:1(2002).


\bibitem{He91a}
M. Herlihy,
Wait-free synchronization.
{\em ACM Trans. Progr. Lang. Syst.}, 13:1(1991),
124--149.

\bibitem{He91} M.P. Herlihy,
Randomized Wait-Free Concurrent Objects,
{\em Proc. 10th ACM Symp. Principles of Distributed Computing}, 1991, 11--21.


\bibitem{HerS93}
 M. Herlihy and N. Shavit,
The topological structure of asynchronous computability,
{\em J. Assoc. Comp. Mach.}, 46:6(1999), 858-923.

\bibitem{HW}
M. Herlihy, J. Wing,
Linearizability: A correctness condition for concurrent
objects, {\em ACM Trans. Program. Languages and Systems},
12(1990), 463--492.


\bibitem{FisLP}
M.J. Fischer, N.A. Lynch, and M.S. Paterson,
Impossibility of Distributed Consensus with One Faulty Processor.
{\em J. Assoc. Comput. Mach.} 32:2(1985), 374--382.



\bibitem{IL87} A. Israeli and M. Li, Bounded Time-Stamps,
{\em Distributed Computing} 6(1993), 205--209.

\bibitem{La86} L. Lamport, On Interprocess Communication Parts I and II,
{\em Distributed Computing} 1(1986), 77--101.

\bibitem{LA-A87} M. Loui, H.H. Abu-Amara,  Memory requirements for
agreement among unreliable asynchronous processes, pp. 163--183 in:
{\em Advances in Computing Research}, Vol. 4, JAI Press, 1987.

\bibitem{LTV89} M. Li, J. Tromp, P.M.B. Vit\'{a}nyi, 
How to share concurrent
wait-free variables,
{\em J. Assoc. Comp. Mach.},   43 (1996), 723-746.

\bibitem{Ly96}
N.A. Lynch, {\em Distributed Algorithms},
Morgan Kaufmann, 1996.


\bibitem{LT87} N.A. Lynch and M. Tuttle, 
An Introduction to Input/Output automata,
{\em CWI-Quarterly}, 2:3(1989), 219--246.


\bibitem{PTTV98}
A. Panconesi, M. Papatrintafilou, P. Tsigas, P. Vitanyi,
Randomized Naming Using Wait-Free Shared Variables,
{\em Distributed Computing}, 11(1998), 113--124.


\bibitem{PF77} G.L. Peterson, M. Fischer, Economical 
solutions for the critical section problem in a distributed
system, {\em Proc. 9th ACM Symp. Theory of Computing}, 1977, 91--97.

\bibitem{Pe83} G.L. Peterson, Concurrent reading while writing,
{\em ACM Trans. Programming Languages and Systems},
5:1(1983), 46--55.

\bibitem{Pl89} S. Plotkin, Sticky bits and universality of consensus,
{\em Proc. 8th ACM Symp. Principles of Distributed Computing},
1989, 159--175.

\bibitem{SinAG94}
A.K. Singh, J.H.  Anderson, and M.G. Gouda,
The Elusive Atomic Register Revisited,
{\em J. Assoc. Comput. Mach.}, 41:2(1994), 311--339.


\bibitem{Sc88} R. Schaffer, {\em On the correctness of atomic multi-writer
registers}, Technical Report MIT/LCS/TM-364, MIT lab. for
Computer Science, June 1988.

\bibitem{SSW90}
M. Saks, N. Shavit, and H. Woll.
Optimal time randomized consensus -- making resilient algorithms fast
in practice,
{\em 2nd ACM Symp. On Discrete Algorithms}, 1991, 351--362.

\bibitem{Ra82} M.O. Rabin,  The choice coordination problem.
{\em Acta Informatica}, 17(1982), 121--134.

\bibitem{TV90}
J.~Tromp and P.~M.~B. Vitanyi,
Randomized wait-free test-and-set,
Manuscript, November 1990.

\bibitem{VA86} P.M.B. Vitanyi, B. Awerbuch,
Atomic Shared Register Access by Asynchronous Hardware, {\em Proc. 27th
IEEE Symp. Foundations of Computer Science}, 1986, 233--243.
(Errata, Ibid.,1987)

\end{thebibliography}
\end{document}